\documentclass{jfm}

\usepackage{subfigure}
\usepackage{graphicx}
\usepackage{natbib}
\usepackage{upmath}
\usepackage{amssymb}
\usepackage{amsbsy}
\usepackage{color}
\usepackage{mathrsfs}

\newcommand\Pec{\mbox{\textit{Pe}}}
\newcommand\Str{\mbox{\textit{Sr}}}
\newcommand\etal{\mbox{\textit{et al.}}}

\newcommand\eg{e.g.\ }
\newcommand\ie{i.e.\ }
\newcommand{\Td}{T_{d}}
\newcommand{\Tu}{T_{u}}
\newcommand{\Th}{T_{h}}
\newcommand{\Tf}{T_f}
\newcommand{\dTdt}{\frac{\partial T}{\partial t}}
\newcommand{\hatdTdt}{\frac{\partial \hat T}{\partial \hat t}}
\newcommand{\dTdx}{\frac{\partial T}{\partial x}}
\newcommand{\hatdTdx}{\frac{\partial \hat T}{\partial \hat x}}
\newcommand{\dTdy}{\frac{\partial T}{\partial y}}
\newcommand{\dTdr}{\frac{\partial T}{\partial r}}
\newcommand{\hatdTdr}{\frac{\partial \hat T}{\partial \hat r}}
\newcommand{\dTdxx}{\frac{\partial^2 T}{\partial x^2}}
\newcommand{\hatdTdxx}{\frac{\partial^2 \hat T}{\partial \hat x^2}}
\newcommand{\dTdyy}{\frac{\partial^2 T}{\partial y^2}}
\newcommand{\hatdTdyy}{\frac{\partial^2 \hat T}{\partial \hat y^2}}

\newcommand{\lak}{\lambda_k}
\newcommand{\lal}{\lambda_l}

\newcommand{\ex}{e^{-x^2}}
\newcommand{\matW}{\mbox{\large $\mathsfbi{W}$}}
\newcommand{\matI}{\mbox{\large $\mathsfbi{I}$}}
\newcommand{\matM}{\mbox{\large $\mathsfbi{M}$}}
\newcommand{\matL}{\mbox{\large $\mathsfbi{\Lambda}$}}

\title[Characterization of a miniature calorimetric flow sensor]{Analytical and experimental characterization of a miniature calorimetric sensor in pulsatile flow}
\author[H. Gelderblom \etal]%
{H.\ns G\ls E\ls L\ls D\ls E\ls R\ls B\ls L\ls O\ls M$^1$%
  \thanks{Present address: Physics of Fluids, Department of Applied Physics, University of Twente, P.O. Box 217, 7500 AE Enschede, The Netherlands. E-mail address for correspondence: h.gelderblom@tnw.utwente.nl},\ns
A.\ns V\ls A\ls N\ns D\ls E\ls R\ns H\ls O\ls R\ls S\ls T$^1$,\ns\break J.\ns R.\ns H\ls A\ls A\ls R\ls T\ls S\ls E\ls N$^2$,\ns M.\ns C.\ns M.\ns R\ls U\ls T\ls T\ls E\ls N$^1$,\ns\break
A.\ns A.\ns F.\ns V\ls A\ls N\ns D\ls E\ns V\ls E\ls N$^3$ \and F.\ns N.\ns V\ls A\ls N\ns D\ls E\ns V\ls O\ls S\ls S\ls E$^1$}

\affiliation{$^1$Department of Biomedical Engineering, Eindhoven University of Technology,
P.O. Box 513, 5600 MB Eindhoven, The Netherlands\\[\affilskip]
$^2$Philips Research Laboratories, High Tech Campus 4, 5656 AE Eindhoven,
The Netherlands\\[\affilskip]
$^3$Department of Mathematics and Computer Science, Eindhoven University of Technology,
P.O. Box 513, 5600 MB Eindhoven, The Netherlands
}

\pubyear{??}
\volume{??}
\pagerange{??--??}
\date{?? and in revised form ??}

\begin{document}
\graphicspath{{figures/}}
\maketitle

\begin{abstract}
The behaviour of a miniature calorimetric sensor, which is under consideration for catheter-based coronary artery flow assessment, is investigated in both steady and pulsatile tube flow. The sensor is composed of a heating element operated at constant power, and two thermopiles that measure flow-induced temperature differences over the sensor surface.

An analytical sensor model is developed, which includes axial heat conduction in the fluid and a simple representation of the solid wall, assuming a quasi-steady sensor response to the pulsatile flow. To reduce the mathematical problem, described by a two-dimensional advection-diffusion equation, a spectral method is applied. A Fourier transform is then used to solve the resulting set of ordinary differential equations and an analytical expression for the fluid temperature is found. To validate the analytical model, experiments with the sensor mounted in a tube have been performed in steady and pulsatile water flow with various amplitudes and Strouhal numbers. Experimental results are generally in good agreement with theory and show a quasi-steady sensor response in the coronary flow regime. The model can therefore be used to optimize the sensor design for coronary flow assessment.

\end{abstract}

\section{Introduction}
\begin{figure}
    \begin{center}
   \includegraphics[width =13cm]{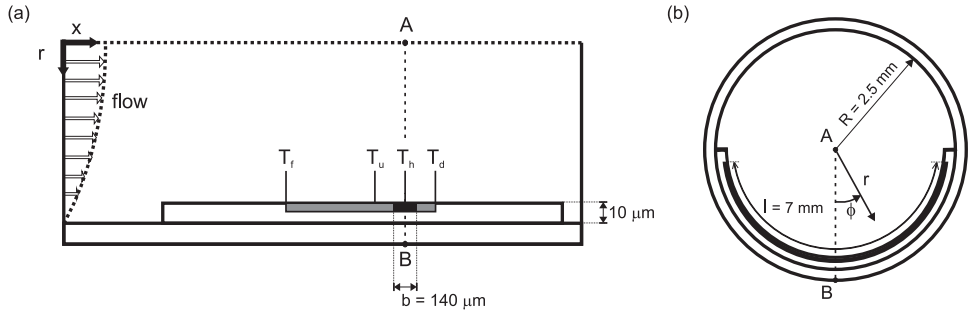}
    \caption{Schematic view of the calorimetric sensor mounted on the inside of the tube wall (not drawn to scale); heater in black, thermopiles in grey, $r$ representing the radial, $x$ the axial, and $\phi$ the circumferential direction. (a) Side-view showing the placement of the sensor elements (heater and thermopiles measuring $\Th-\Tf$ and $\Td-\Tu$) with respect to the fluid flow. (b) Cross-section showing the positioning of the flexible device inside the tube.}\label{sensorlayout}
    \end{center}
\end{figure}

Flow sensors based on forced convective heat-transfer, such as hot-film anemometers, can be used for the assessment of arterial blood flow \cite[][]{Seed70, Clark74, Nerem76}. In a recent study, \cite{Tonino09} showed that if the treatment of patients with coronary artery disease is based on an indirect measure for coronary flow (derived from coronary pressure measurements), the clinical outcome improves significantly. Clearly, direct flow assessment by miniature sensors that can be introduced into the coronary arteries would provide even more information about the condition of these arteries \cite[][]{Veer09}. In this study, we aim to characterize the behaviour of such a miniature convective heat-transfer sensor in steady and pulsatile tube flow, through both an analytical and an experimental approach. The sensor is based on a calorimetric flow measurement principle: it consists of a small aluminium heating element of width $b$=140 $\umu$m, operated at constant power, and two polysilicon thermopiles that measure flow-induced temperature differences over the sensor surface. These sensor elements are embedded in a flexible polyimide substrate having a thickness of 10 $\umu$m; see figure \ref{sensorlayout}\,(\textit{a}). In order to use it for coronary flow assessment, the flexible device is bent around a catheter guide wire, which can be inserted into the coronary arteries. In our characterization study however, the device is mounted at the inner wall of a tube, to be able to subject it to a well-defined flow regime. The length $l$ of the device is equal to approximately half the circumference of the inner tube wall; see figure \ref{sensorlayout}\,(\textit{b}). The temperature difference between two positions 100 $\umu$m downstream and 100 $\umu$m upstream from the heater centre ($\Td-\Tu$) is measured, as well as the heater temperature $\Th$ with respect to the ambient fluid temperature $\Tf$ far upstream (2000 $\umu$m from the heater centre). In absence of flow, heat transfer from the sensor to the fluid occurs solely through conduction, resulting in a symmetric temperature distribution over the sensor surface. If a certain fluid flow exists, the advective heat-transfer leads to an asymmetric temperature distribution. The resulting temperature differences are a measure for the flow \cite[][]{Elwenspoek99}. Flow reversal will lead to a sign change in $\Td-\Tu$, and hence can be detected, which is an advantage compared to the conventional hot-film anemometers \cite[][]{Oudheusden89}.

Two important dimensionless parameters that appear in the study of thermal sensors in time-dependent flow are the P\'eclet number $\Pec$, and the Strouhal number $\Str$. Here, $\Pec$ is a measure for the importance of  advective compared to conductive heat-transfer, $\Str$ for the importance of unsteady compared to advective temperature variations. Formal expressions for $\Str$ and $\Pec$ are given further on (see (\ref{str}) and (\ref{pec}), respectively).

Many experimental and analytical studies of hot-film anemometers have been reported in the literature. Experiments with these kinds of probes have been performed by, among others, \cite{Seed70}, \cite{Clark74}, \cite{Ackerberg78}, and \cite{Steenhoven91}. \cite{Liepmann54} derived a theoretical relation between the amount of convective heat loss from a hot surface and the local steady wall-shear rate. \cite{Pedley72,Pedley76} and \cite{Menendez85} extended this work to include unsteady, pulsatile flows.

For miniature flow sensors like the one presented here, the theory developed for hot-film anemometers is not applicable. First of all, because in a thermal sensor with small dimensions, generally operated at small $\Pec$, heat conduction in the flow direction cannot be neglected, as is done in the usual boundary-layer approximation (see \eg \cite{Liepmann54, Pedley72, Menendez85}). In numerical studies, \cite{Tardu05} and \cite{Rebay07} showed that this axial conduction has a considerable influence on the response of small hot-film gauges. \cite{Ackerberg78} derived an analytical solution for the heat transfer from a finite strip for $\Pec\rightarrow 0$, in steady flow. \cite{Ma93} examined the leading and trailing edge of a micro-sensor in steady flow separately to obtain an analytical solution for the entire sensor surface, in analogy to the method used by \cite{Springer73} and \cite{Springer74}.

Second, most analytical studies consider the problem of a uniform surface temperature on the heated element, while our sensor is operated at constant power, which is better described by a heat-flux boundary condition. \cite{Liu94} and \cite{Rebay07} used a constant heat-flux boundary condition on the surface of a heated element embedded in an adiabatic wall. However, the thermopiles of our sensor consist of conductive polysilicon, and therefore heat will not only be transferred from the heater to the fluid directly, but also via the surrounding material.
In that case, we end up with a conjugate heat-transfer problem, where the heat source is known, but the interface temperature and heat flux to the fluid are unknown; \cite{Tardu05} studied this problem numerically. \cite{Stein02} derived an analytical solution downstream of a flush-mounted heat source in steady flow. \cite{Cole08} considered conjugate heat-transfer from a steady-periodic heated film, also in steady flow. To the authors' knowledge, only a few analytical models specific for calorimetric sensors exist.
\cite{Lammerink93} described an experimental study and a relatively simple analytical model of such a sensor, but in steady flow and with a different sensor geometry. Calorimetric sensors on highly conductive silicon wafers are described by \cite{Oudheusden91}, also in steady flow and with very small on-sensor temperature differences compared to the sensor overheat. Experimental studies with calorimetric flow sensors in steady flow have been reported by \cite{Lammerink93} and \cite{Nguyen95}. However, no data on unsteady flow experiments with this type of sensor are available.

Our analysis of the calorimetric flow sensor focuses on the derivation of a new analytical model for the temperature distribution in a pulsatile fluid flow over a small heated element operated at constant power. Experiments with the sensor in steady and pulsatile water flow with Strouhal numbers and amplitudes in the expected physiological flow range are carried out to verify our theoretical predictions.
In \S\ref{mat1} the mathematical formulation of our problem is given in terms of a two-dimensional advection-diffusion equation. We circumvent the coupling of the heat-transfer problems in the fluid and the substrate by approximating the heat flux from the sensor to the fluid, and use this approximation as a boundary condition for the fluid compartment. The axial conduction term is retained, and therefore our solution holds for all $\Pec$-values. Since the heat flux at the boundary is approximated by a continuous function, the leading and trailing edge of the heater do not have to be treated separately \cite[][]{Ma93}, resulting in one solution for the complete domain.  When applied in coronary flow, our sensor will be operated at small Strouhal numbers, therefore a quasi-steady sensor response to the pulsatile flow is assumed. As described in \S\ref{mat2}, a spectral method is applied to reduce the mathematical problem to one dimension. Then, a Fourier transform is used to solve the resulting set of ordinary differential equations. The experimental technique is described in \S\ref{experiments}. In \S\ref{results}, the experimental results are compared to the theoretical predictions, and found to be in good agreement. The model developed not only leads to theoretical understanding of the operating principle of the sensor, but it can also be used to optimize the sensor design, as is demonstrated in \S\ref{results}.

\section{Mathematical problem formulation}\label{mat1}
In order to formulate an analytical model for the sensor in a pulsatile tube flow (see figure \ref{sensorlayout}), a cylindrical coordinate system ($r,\phi,x$) is adopted, where the main flow is in axial or $x$-direction, $r$ is the radial, and $\phi$ the circumferential coordinate. The origin of this system is chosen such that $x=0$ at the heater centre. The pulsatile fluid flow is assumed to be fully developed, and, because the temperature difference between the heater and the oncoming fluid is relatively small, temperature independent. The typical buoyancy-driven radial velocity can be estimated from the momentum equation in the radial direction using the Boussinesq approximation. For our configuration, the ratio of radial to axial velocity is of order $10^{-2}$, hence free convection can be neglected.

The basic problem is thus reduced to that of finding the temperature distribution $T(x,r,\phi,t)$, with $t$ the time, in a prescribed pulsatile fluid flow in a tube of radius $R$, which is heated by a time-constant prescribed heat influx in a small region of length $l$ and width $b$ around the tube wall; the remaining part of the wall is thermally insulated.
The equation governing the temperature distribution in the fluid is the thermal energy equation in the tube ($x\in\mathbb{R}, 0\leq r\leq R$, $-\pi\leq \phi\leq \pi$)
\begin{equation}
\dTdt+u(r,t)\dTdx= \alpha\left[\dTdxx+\frac{1}{r}{\frac{\partial}{\partial r}}\left(r\dTdr\right)+\frac{1}{r^2}\frac{\partial^2T}{\partial \phi^2}\right]\label{energyeq4},
\end{equation}
together with the boundary condition at the tube wall, $r=R$, describing the prescribed heat influx,
\begin{eqnarray}\label{influx}
k\dTdr(x,R,\phi,t)&=&q(x),~\mathrm{if}~\frac{-l}{2R}<\phi<\frac{l}{2R},\\
&=&0,~\mathrm{if}~|\phi|>\frac{l}{2R}.\nonumber
\end{eqnarray}
Here, $u$ is the velocity in $x$-direction, $\alpha$ the thermal diffusivity, and $k$ the thermal conductivity. Heat influx $q$ is in Wm$^{-2}$, such that the power supplied to the heater in W is given by
\begin{equation}
Q=l\int_{-\infty}^{\infty} q(x)dx.\label{heaterpower}
\end{equation}
Considering the case that $u>0$, \ie the fluid is flowing in the positive $x$-direction, we state that $T$ must go to $T_o$, the initial fluid temperature, for $x\rightarrow -\infty$, but that $T$ for $x\rightarrow +\infty$ must tend to a value $T_{\infty}>T_o$ for a quasi-steady solution to exist. In that case the total heat-transfer rate $Q$ into the fluid is balanced by the advective heat outflow in positive $x$-direction, equal to $\rho cD(T_{\infty}-T_o)$, with $\rho$ the density, $c$ the specific heat, and $D=2\pi\int_0^R u(r,t)rdr$, the total volumetric flow rate at time $t$.

To emphasize the effect of the two different length scales that arise in the problem, \ie heater width $b$ and tube radius $R$, we introduce the following dimensionless variables
\begin{equation}
\hat x=\frac{x}{b},\quad\hat r=\frac{r}{R},\quad\hat u=\frac{u}{V},\quad\hat t=\frac{t}{t_c},\quad\hat T=\frac{T-T_o}{T_c}, \label{dimlqu}
\end{equation}
with $V$ the typical axial velocity, $t_c$ the time scale for temperature variations, and $T_c$ the typical temperature scale. The characteristic parameter values can be found in table \ref{tabparan}, appropriate choices for $t_c$ and $T_c$ are explained below. Note that the heater width is used as the characteristic length scale in $x$-direction, implying that we will look for changes in temperature $T$ in the direct axial vicinity of the heater, which is where $\Td$ and $\Tu$ are measured.
\begin{table}
    \begin{center}
    \begin{minipage}{10cm}
        \begin{tabular}{@{}cccl@{}}
        {Parameter} & {value}& {unit}&{description}\\[3pt]
       $T_o$&20&$^\circ$C&outer flow temperature\\
       $T_c$&11.7&$^\circ$C&temperature scale\\
       $Q$&80&mW&heater power\\
       $x_h$&0&$\umu$m&heater centre\\
       $x_d$&$100$&$\umu$m&position where $\Td$ is measured\\
       $x_u$&$-100$&$\umu$m&position where $\Tu$ is measured\\
       $x_f$&$-2000$&$\umu$m&position where $\Tf$ is measured\\
       $b$&$140$&$\umu$m&heater width\\
       $l$&$7000$&$\umu$m&heater length\\
       $\sigma$&$70$&$\umu$m&standard deviation of the assumed\\
       &&&boundary heat-flux distribution\\
       $R$&2.5&mm& inner tube radius\\
       $V$&$0.1$&ms$^{-1}$&typical axial velocity\\
       $S_0$&115&s$^{-1}$&mean wall-shear rate\\
       $\alpha$&$1.44\cdot 10^{-7}$&m$^2$s$^{-1}$&thermal diffusivity \footnote[2]{\cite[see][p.~860]{Incropera07}}\\
       $k$&$0.606$&Wm$^{-1}$K$^{-1}$&thermal conductivity\footnotemark[2]\\
       $\nu$&$1\cdot10^{-6}$&m$^2$s$^{-1}$&kinematic viscosity\footnotemark[2]\\
       $\omega$&$2\pi$&rad s$^{-1}$&angular frequency\\
       \end{tabular}
       \end{minipage}
    \end{center}
  \caption{The parameter values used in the analytical model, based on the experimental set-up.} \label{tabparan}
\end{table}
By substituting (\ref{dimlqu}) into (\ref{energyeq4}), we obtain  
\begin{equation}
\frac{b}{t_c V}\hatdTdt+\hat u(\hat r,\hat t)\hatdTdx= \frac{\alpha}{b V}\hatdTdxx+\epsilon^2\left[\frac{1}{\hat r}\frac{\partial}{\partial \hat r}\left(\hat r\hatdTdr\right)+\frac{1}{\hat{r}^2}\frac{\partial^2 \hat{T}}{\partial \phi^2}\right],\label{energyeq5a}
\end{equation}
with $\epsilon=\sqrt{\left(\alpha b/V\right)}/R=0.006 \ll 1$.
Since, in (\ref{energyeq5a}), the small number $\epsilon^2$ appears in front of the highest derivative with respect to $\hat {r}$, one can expect a boundary layer to develop at the tube wall; \ie at $\hat {r}=1$. The outer solution at leading order, with $\epsilon=0$, is the trivial solution $\hat T=0$. The temperature problem is thus confined to a small region close to the sensor surface: the thermal boundary layer of thickness $\delta_T$. 

Our sensor measures the temperature difference $\Td-\Tu$ at distances in the order of magnitude of $b$ up- and downstream of the heater centre for $|\phi|<l/2R$. Hence, in the region of interest for our sensor $|x|=O(b)$ and $\delta_T\ll R$. Within this region, the problem is independent of the $\phi$-coordinate. The characteristic length-scale for conduction in the $\phi$-direction, heater length $l=O(R)$, is much larger than the length scale for conduction in the axial direction, heater width $b$. For our sensor $b/l=0.02$; hence conduction in the $\phi$-direction and edge effects occurring at $\phi=\pm l/2R$ can be neglected.

\begin{figure}
\begin{center}
   \includegraphics[width =10cm]{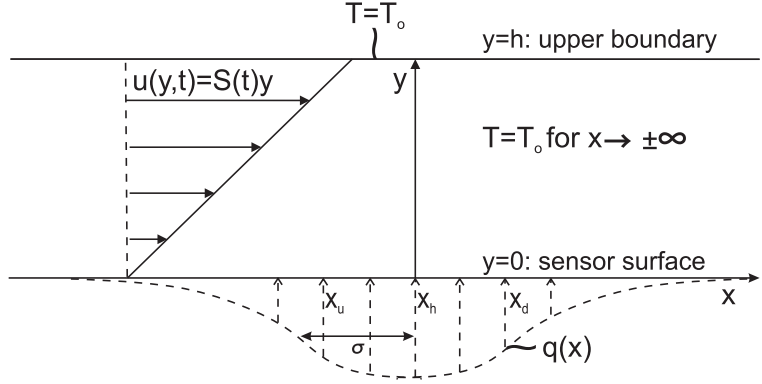}
   \caption{Scheme of the problem geometry.}\label{figanalmod}
   \end{center}
\end{figure}

Since the thermal boundary layer thickness $\delta_T$ is much smaller than the tube radius, the tube wall in a $b$-environment of the heater can be considered flat. We therefore adopt a spatial rectilinear coordinate system ($\hat x,\hat y$), where $\hat x$ is the surface coordinate in the flow direction, and $\hat y$ is the stretched coordinate normal to the surface, defined as $\hat y=(R/\delta_T)(1-\hat r)$; see figure \ref{figanalmod}. As a further approximation, we confine the domain for the inner solution to a strip of finite height $h$. At the upper boundary of the strip, $y=h$ (with $y=\delta_T\hat{y}$), we then require that $\hat{T}=0$, to match the inner solution to the outer one; see figure \ref{figanalmod}. How to choose $h$ such that the solution in a $b$-environment of the heater, where the thermal boundary layer is still thin, is not influenced by the finite size of the domain in $y$-direction will be explained further on in this section; see (\ref{eq2.2}).

Further downstream (for $x>b$, hence outside our region of interest) the thermal boundary layer widens, due to radial conduction. Both curvature and $\phi$-dependence will enter the problem again, while axial conduction will become negligible. Even further downstream ($\hat{x}>R/b$), the fluid temperature will become uniform in each cross-section, with $T\to T_{\infty}=T_0+Q/\rho cD$. It is therefore important to note that, given the simplifications described above, our method will only yield the correct solution in an $b$-environment of the heater, \ie the region of interest for our sensor.

As a further approximation, we assume that the wall-shear rate is the only flow parameter that influences the heat transfer from the sensor surface, implying that the velocity profile may be approximated linearly throughout the thermal boundary layer \cite[see][]{Pedley72}. Since our domain is now restricted to a strip of finite height $h$, the linearization of the velocity profile is valid throughout the complete domain. This approximation requires the Stokes layer thickness $\delta_S$ to be much larger than the thermal boundary layer thickness $\delta_T$. In that case the velocity $u$ within the thermal boundary layer can be approximated by ($y=R-r=\delta_T \hat{y}$)
\[u(y,t)=\left.\frac{\partial u}{\partial y}\right|_{y=0}~y=S(t)y,\]
with $S$ the wall-shear rate, which is, in a pulsatile tube flow, given by
\[S(t)=S_0\left[1+\beta \sin\left(\omega t\right)\right],\]
with $S_0$ the mean, $\omega=2\pi f$ the angular frequency, and $\beta$ the amplitude of the shear-rate oscillations. This implies that when $\beta>1$, backflow is involved. Since for coronary flow, the order of magnitude of $\beta$ will be about one, the dimensionless shear rate $\hat S(\hat t)=S(t)/S_0$ is an $O(1)$-function of $t$. Hence, we have
\[ u(y,t)=S(t)y=S_0 \delta_T \hat y \hat S(\hat t),  \]
yielding
\[ V= \delta_T S_0,\qquad\mathrm{and}\qquad\hat u(\hat y, \hat t) = \hat S(\hat t)\hat y. \]

To find an expression for the thermal boundary layer thickness $\delta_T$, we write (\ref{energyeq5a}) in terms of $\hat y,~ \hat u(\hat y, \hat t)$, and $\hat S(\hat t)$ as
\begin{equation}
\frac{b}{t_c S_0\delta_T}\hatdTdt+\hat S(\hat t)\hat y\hatdTdx= \frac{\alpha}{b S_0\delta_T}\hatdTdxx+\frac{\alpha b}{\delta_T^3 S_0}\hatdTdyy.\label{energyeq5b}
\end{equation}
When advection in $x$- and conduction in $y$-direction are the two dominant effects, $\alpha b /S_0\delta_T^3$ must be of $O(1)$, and hence the thermal boundary layer thickness is given by \cite[see][]{Liepmann54}:
\begin{equation}
\delta_{T}= \left(\alpha b/S_0\right)^{1/3}. \label{thermalbl}
\end{equation}
The linearization of the velocity profile within the thermal boundary layer is allowed if $\delta_T \ll\delta_S.$
The Stokes layer thickness in a fully developed pulsatile tube flow is given by \cite[p. 367]{Schlichting00}
\[\delta_S=\left(\nu/\omega\right)^{1/2},\]
with $\nu$ the kinematic viscosity. For a 1 Hz pulsatile water flow we get $\delta_S=4\cdot10^{-4}$ m.
The requirement $\delta_T \ll\delta_S$ leads us to an estimate for the admissible shear rate:
\[S_0\gg \alpha b/\delta_S^3=0.3~\mathrm{s}^{-1}.  \]
This requirement is amply satisfied, since our experiments are performed at a mean wall-shear rate of about 115 s$^{-1}$. Furthermore, $\delta_T$ and $b$ are of the same order of magnitude in this range of shear rates, allowing the use of $b$ as length scale in both $x$- and $y$-direction. This already indicates that the axial conduction term in (\ref{energyeq5b}) can certainly not be neglected within the region of interest for our sensor. This is further confirmed by the magnitude of the coefficient of $x$-conduction; $\alpha/b S_0\delta_T=0.16$  (see table \ref{tabparan} for the parameter values used).

The magnitude of the dimensionless group in front of the unsteady term in (\ref{energyeq5b}), the Strouhal number $\Str=b/t_cS_0 \delta_T \approx 1/t_cS_0$, if $b/\delta_T \approx 1$,
depends on the choice of the characteristic time scale $t_c$. The goal of this study is not to analyse start-up processes that occur when switching on the heater, but to describe the periodic variations in the sensor response. Therefore, the oscillation time, $1/\omega$, is used as characteristic time scale; hence 
\begin{equation}
\Str =\omega/S_0.\label{str}
\end{equation} 
Eventually, this sensor will be used for coronary flow measurements, where the estimated mean shear rate the sensor experiences when positioned on a guide wire is in the order of magnitude of 1000 s$^{-1}$. Hence, $\Str$ is generally small for our ultimate application (typically $\Str< 0.1$, assuming measurement of the first ten harmonics is sufficient for reconstruction of the coronary flow signal \cite[][p. 157]{Milnor89}). We therefore assume the fluid temperature distribution to be quasi-steady, thereby neglecting the unsteady term in the thermal energy equation. In the quasi-steady approximation time $t$ represents a parameter rather than a variable. From here on, we therefore omit the explicit dependence on $\hat t$ ($\hat{T}=\hat{T}(\hat x,\hat y)$); in fact, the role of $\hat t$ is now taken over by the shear rate $\hat S$ ($\hat S = 1+\beta \sin{(\hat t)}$).

Another simplification is that heat loss through the insulating back of the tube in which the sensor is mounted, is neglected: all heat produced by the heater is assumed to be transferred to the fluid. Capacitive effects, which may cause the heat transfer to the fluid to vary in time, are also neglected. According to \cite{Tardu05} this is reasonable, since the thermal diffusivities of the sensor components are two orders of magnitude higher than that of water. Since the exact shape of the heat-flux distribution from the sensor substrate to the fluid depends on the temperature distribution in the fluid, this leads to a conjugate heat-transfer problem, which is hard to solve. We circumvent this coupling of the fluid and substrate temperatures by a much simpler approach: we approximate the shape of the heat-flux distribution from substrate to fluid and use this as a boundary condition for the fluid problem. If all the heat would be transferred from the heater to the fluid directly, \ie when there is perfect insulation outside the heater compartment, a rectangular-shaped heat-flux boundary condition would be most realistic. For our sensor, however, conduction of heat from the heater towards the other sensor components will smooth the rectangular-shape, leading to a more Gaussian-shaped heat-flux profile, with some deviations due to the asymmetric temperature distribution in the fluid. As a simple approximation of the real heat-flux boundary condition, we therefore use a Gaussian distribution with a standard deviation $\sigma$ equal to half the heater width, hence $\sigma=b/2$:
\begin{equation}
q(x)=\left. -k\dTdy \right|_{y=0}=\frac{Q}{l\sigma \sqrt{2\pi}}e^{-\frac{(x-x_h)^2}{2\sigma^2}}\label{heatsource},
\end{equation}
with $x_h$ the position of the heater centre, and $Q/l$ the total amount of heat transferred from the sensor to the fluid, per unit of length in $z$-direction (in Wm$^{-1}$), as given by (\ref{heaterpower}).

The resulting dimensional thermal energy equation and boundary conditions describing the quasi-steady problem for the fluid temperature $T=T(x,y)$ within the thermal boundary layer or strip are given by:
\begin{eqnarray}
&S&y\dTdx = \alpha \left(\dTdxx+\dTdyy\right),\qquad x \in \mathbb{R},~0\leq y \leq h,  \nonumber\\
&T&(\pm \infty,y)=T_0,\quad\dTdy(x,0) =-\frac{Q}{k l\sigma \sqrt{2\pi}}e^{-\frac{(x-x_h)^2}{2\sigma^2}},\quad T(x,h)=T_0.
\label{eq2.1}
\end{eqnarray}
Figure \ref{figanalmod} shows a schematic view of the resulting problem to be solved.
Note that, although we are only interested in the solution close to the heater (for $|x|=O(b)$), we have, for mathematical ease, extended the domain in $x$-direction to infinity. 
The outflow boundary condition used, implies that in our model, all heat will eventually escape through the upper boundary $y=h$. Since, for the correct choice of $h$, this happens sufficiently far away from the heater, it does not influence the solution in a $b$-environment of the heater. To ensure this, the upper boundary of the domain has to be located sufficiently far outside the thermal boundary layer for $x=O(b)$. Therefore, $h$ is taken equal to $n$ times the estimated thermal boundary layer thickness, where $n=1,2,3,...$. Further on, it will be shown that for $n \geq 4$ the solution becomes independent of $n$.
Since the thermal boundary layer thickness $\delta_T$ according to (\ref{thermalbl}), but with $S_0$ replaced by $S$, depends on the actual wall-shear rate, $h$ depends on $S$ as well. This motivates us to choose $h$ as (note that in the following we use $\sigma \sqrt{2}$ instead of $b$ as characteristic unit of length)
\begin{equation}
h=n\left(\frac{\alpha \sigma\sqrt{2}}{|S|}\right)^{1/3}.
\label{eq2.2}
\end{equation}
Hence, in our quasi-steady approximation we solve the problem for each value of $S$ separately, choosing the upper boundary accordingly. The advantage of this $S$-dependent position of the upper boundary will become clear in the next section. We note that $h$ can become large, \ie larger than $b$, for small values of $|S|$, specifically for $|S|<0.3$ s$^{-1}$. Then, the thermal energy equation is no longer advection-, but diffusion-dominated and in that case $\delta_T$ must be taken equal to $b$. However, for our problem this happens in a very short period of time (less than 0.1\% of one period of $S(t)$) and it is therefore not relevant for our solution. 
 
We introduce a new scaling by using the dimensionless variables and parameters:
\begin{eqnarray}
\tilde{x}&=&\frac{x-x_h}{\sigma\sqrt{2}},\quad\tilde{y}=\frac{y}{\sigma\sqrt{2}},\quad
\tilde{T}(\tilde{x},\tilde{y})=\frac{T(x,y)-T_o}{T_c},\nonumber\\
T_c&=&\frac{Q}{k l\sqrt{\pi}},\quad \tilde{h}=\frac{h}{\sigma\sqrt{2}},\quad\tilde{\alpha}= \frac{\alpha}{2\sigma^2 S},
\label{eq1.4}
\end{eqnarray}
where the temperature scale $T_c$ is based on the heat source term (\ref{heatsource}). We define the P\'eclet number as  
\begin{equation}
Pe=\frac{2\sigma^2S}{\alpha},\label{pec}
\end{equation} 
hence $\tilde{\alpha}$=$1/\Pec$. 

Omitting the tildes, the newly scaled system for $T=T(x,y)$ reads
\begin{eqnarray}
y\dTdx &=& \alpha \left(\dTdxx+\dTdyy\right),\quad x \in \mathbb{R},~0\leq y\leq h,  \nonumber \\
T(\pm \infty,y)&=& 0,\quad\dTdy(x,0) =-e^{-x^2},\quad T(x,h)=0,
\label{eq2.3}
\end{eqnarray}
where
\begin{equation}
h=n\alpha^{1/3},
\label{eq2.4}
\end{equation}
with $n$ still to be chosen. Hence, $T(x,y)$ depends on only two parameters, $\alpha$ and $n$: $T(x,y)=T(x,y;\alpha,n)$. However, if $n$ is taken sufficiently large, \ie $n\ge 4$, then solution $T$ in a $b$-environment of the heater becomes independent of $n$, and $\alpha$ is the only parameter remaining.

\section{Analytical solution method}\label{mat2}
To solve the system (\ref{eq2.3}) a spectral method is used, which reduces the partial differential equation in (\ref{eq2.3}) to a set of ordinary differential equations. To apply this method, we first make the boundary conditions homogeneous, by writing
\begin{equation}
T(x,y)= (h-y)e^{-x^2} + T_1(x,y),
\label{eq2.5}
\end{equation}
leaving for $T_1$ the equation
\begin{equation}
y\frac{\partial T_1}{\partial x} - \alpha\left(\frac{\partial^2 T_1}{\partial x^2}+ \frac{\partial^2 T_1}{\partial y^2}\right)=R(x,y)~,
\label{eq2.6}
\end{equation}
with homogeneous boundary conditions and with
\begin{equation}
R(x,y)=\left(h-y\right)\ex\left[2xy+\alpha\left(4x^2-2\right)\right].
\label{eq2.7}
\end{equation}
For the spectral method, we introduce the trial functions $v_k(y)$, given by
\begin{equation}
\frac{\mathrm{d}^2 v_k}{\mathrm{d}y^2} = -\lambda_k^2 v_k,\quad\frac{\mathrm{d} v_k}{\mathrm{d}y}(0) =0,\quad v_k(h) =0,
\label{eq2.8}
\end{equation}
yielding
\begin{equation}
v_k(y) = \cos(\lambda_k y),\quad\lambda_k = \frac{(2k-1)\pi}{2h},\quad k=1,2,....
\label{eq2.9}
\end{equation}
Here we see the advantage of truncating the infinite half-space to a strip of finite height. 
Next, we decompose $T_1$ into a linear combination of the trial functions $v_k$ according to
\begin{equation}
T_1(x,y)= \sum_{k=1}^\infty C_k(x)v_k(y) \approx \sum_{k=1}^K C_k(x)v_k(y),
\label{sol}
\end{equation}
where in the last step we have truncated the series after $K$ terms (as demonstrated in \S\ref{results}, $K=5$ is more than sufficient for obtaining precise numerical results when the height of the strip is chosen according to the actual wall-shear rate).
Substituting (\ref{eq2.5}) and (\ref{sol}) into (\ref{eq2.3}), we obtain
\begin{equation}
\sum_{l=1}^K \left[y\frac{\mathrm{d} C_l}{\mathrm{d} x}
 - \alpha\left(\frac{\mathrm{d}^2 C_l}{\mathrm{d} x^2}- \lal^2 C_l\right)\right]v_l(y)
=R(x,y).
\label{eq2.10}
\end{equation}
Taking the inner product of (\ref{eq2.10}) with functions $v_k(y)$, with the inner product of a function $u$ with $v$ defined as
\[(u,v)\equiv \int_0^{h} u(y)v(y) \mathrm{d}y,\]
we arrive at an equation for the array $\boldsymbol{C}$, consisting of $K$ elements $C_k$,
\begin{equation}
h^2 \matW \frac{\mathrm{d} \boldsymbol{C}}{\mathrm{d} x}
 - \alpha\frac{h}{2}\left( \frac{\mathrm{d}^2 \boldsymbol{C}}{\mathrm{d} x^2} - \matL \boldsymbol{C}\right)
=\boldsymbol{R},
\label{energyeq6}
\end{equation}
with $\matW$ a $K \times K$-matrix with elements $W_{kl}$, given by (\ref{Wkl}), $\matL$ a $K \times K$ diagonal matrix with elements $\Lambda_{kk}=\lak^2$ and $\boldsymbol{R}$ a $K$-array with elements $R_k$,  given by (\ref{Rk}):
\begin{subeqnarray}
W_{kl}&=&\frac{1}{h^2}\int_0^{h} yv_k(y)v_l(y)dy = \int_0^{1} \hat yv_k(h\hat y)v_l(h\hat y)d\hat y,
\slabel{Wkl}\\
R_k(x)&=&\int_0^{h}R(x,y)v_k(y)dy.
\slabel{Rk}
\end{subeqnarray}

We introduce the Fourier transform of $\boldsymbol{C}(x)$ by
\begin{equation}
\boldsymbol{c}(\zeta) = \frac{1}{\sqrt{2\pi}}~\int_{-\infty}^{\infty} \boldsymbol{C}(x)e^{-i\zeta x} dx =\mathscr{F}\{\boldsymbol{C};\zeta \}.
\label{eq2.11}
\end{equation}
By taking the Fourier transform of (\ref{energyeq6}), after dividing it by $\alpha h/2$, we obtain the algebraic equation for $\boldsymbol{c}$:
\begin{equation}
\left(\zeta^2 \matI+ i\zeta \frac{2h}{\alpha} \matW + \matL \right)\boldsymbol{c}(\zeta) = \matM(\zeta) \boldsymbol{c}(\zeta) = \boldsymbol{r}(\zeta),
\label{eq2.12}
\end{equation}
with $\matI$ the unity $K \times K$-matrix, and
\begin{subeqnarray}
\matM(\zeta)&=&\zeta^2 \matI+ i\zeta \frac{2h}{\alpha} \matW + \matL,
\slabel{Mkl}\\
\boldsymbol{r}(\zeta)&=&\mathscr{F}\{\frac{2}{\alpha h}\boldsymbol{R};\zeta \}.
\slabel{rk}
\end{subeqnarray}
We can, using Mathematica 6 (Wolfram Research, Champaign, USA), invert the $K \times K$-matrix $\matM$ analytically, by which we find
\begin{equation}
\boldsymbol{c}(\zeta) = \matM^{-1}(\zeta)\boldsymbol{r}(\zeta),
\label{eq2.13}
\end{equation}
and by taking the inverse Fourier transform of this result, we obtain the solution for the array $\boldsymbol{C}(x)$ as
\begin{equation}
\boldsymbol{C}(x) = \frac{1}{\sqrt{2\pi}}~\int_{-\infty}^{\infty} \boldsymbol{c}(\zeta)e^{i\zeta x} d\zeta =
  \frac{1}{\sqrt{2\pi}}~\int_{-\infty}^{\infty} \matM^{-1}(\zeta) \boldsymbol{r}(\zeta)e^{i\zeta x} d\zeta~.
\label{eq2.14}
\end{equation}
The latter integral is evaluated numerically using Mathematica~6. The temperature $T$ is now determined by (\ref{eq2.5}) and (\ref{sol}) with $K=5$ and $n=4$.

\section{Experimental methods}\label{experiments}
In the experimental set-up, the device was mounted to the inner wall of a tube with an inner diameter of 5 mm (see figure \ref{sensorlayout}\,(\textit{b})). The tube was made of PMMA, which is an insulating material, to prevent heat loss through the back of the device. The device covered about half of the tube perimeter. The polyimide foil including the sensor components has a thickness of only 10 $\umu$m (see figure \ref{sensorlayout}\,(\textit{a})), and since the very small step it causes in the tube wall is located about 2 mm away from the actual sensor components (on both sides) this does not significantly disturb the flow pattern near the sensor. In all experiments, the heater was supplied with a power of 80 mW by a voltage source (EST 150, Delta Elektronika, Zierikzee, The Netherlands). Two multimeters (DMM 2000, Keithley Instruments Inc, Cleveland, USA) were used to register the output of the thermopiles that measure $\Td-\Tu$ and $\Th-\Tf$.

To ensure fully developed flow over the sensor, the measurement section was located 112 tube diameters from the tube entrance, which is, even at the highest Reynolds number reached ($\approx 500$), well beyond the laminar entrance region. As test fluid, tap water at room temperature was used.
Steady flow through the set-up was generated by a stationary pump (Libel-Project, Alkmaar, The Netherlands). The amount of flow could be adjusted using a clamp. The oscillatory component was added to the mean flow by a piston pump, driven by a computer-controlled motor (ETB32, Parker Hannifin, Offenburg, Germany). Downstream of the sensor, the flow was registered by an ultrasonic flow probe (4PSB transit time perivascular probe, Transonic Systems Inc, Ithaca, USA), which was used as a reference. The signals from the flow probe and the multimeters were recorded simultaneously and transferred to a computer via an acquisition board with a sampling frequency of 20 Hz.

The output voltage of a thermopile is proportional to the temperature difference between its ends via the Seebeck coefficient, which depends on the composition of the thermocouple leads \cite[see][]{Herwaarden89}. By scaling the stationary sensor response $\Th-\Tf$ at $S=115$ s$^{-1}$ to the analytical value at this shear rate, we found the Seebeck coefficient for the thermocouples in our sensor to be 305 $\umu$VK$^{-1}$. This Seebeck coefficient, which is only a scaling value for the experimental data and does not influence the shape of the responses, is used for all experimental results shown in \S\ref{results}. 

Experiments under both steady and unsteady flow conditions were performed. In steady flow, the P\'eclet number (see (\ref{pec})) was varied from 0 (at zero flow; $S=0$ s$^{-1}$) to 34 (at a flow of 368 ml min$^{-1}$; $S=500$ s$^{-1}$). The wall-shear rate at the sensor surface was calculated from the flow measured by the ultrasonic probe assuming a Poiseuille velocity profile.

For the unsteady case the Strouhal number (see (\ref{str})) was varied from 0.01 to 0.1 by varying the oscillation frequency from 0.2 to 2 Hz, and the amplitude ($\beta$) was varied between 0.8 and 1.2 (corresponding to the expected coronary flow regime), keeping the mean shear rate constant at about 115 s$^{-1}$. In unsteady flow, the wall-shear rate was derived from the flow measurements assuming a Womersley velocity profile \cite[][]{Womersley55}.

\section{Results and discussion}\label{results}

\begin{figure}
\begin{center}
         \includegraphics[width=13cm]{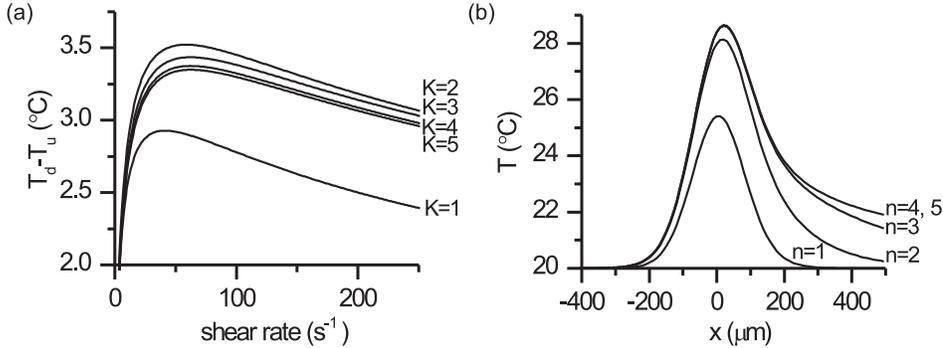}
   \caption{The influence of (a) the number of cosine-terms $K$ on the solution $\Td-\Tu$ and (b) parameter $n$ on the solution $T$ at $y=0$, $S=100$ s$^{-1}$.}\label{convergence}
    \end{center}
\end{figure}

The cosine-series used in (\ref{sol}) converged quite rapidly: only five terms sufficed for an accurate approximation of the solution; see figure \ref{convergence}\,(\textit{a}). The rapid convergence is a consequence of the dependence of $h$ on the actual wall-shear rate $S$; if $h$ would have been fixed for all $S$, it could become much larger than the boundary-layer thickness, leading to slow convergence of the cosine series.
The parameter $n$ (see (\ref{eq2.4})) was chosen such that the position of the boundary condition did not influence the solution at the sensor surface in a $b$-environment of the heater; $n=4$ was found to be large enough to ensure this, as demonstrated in figure \ref{convergence}\,(\textit{b}).

\begin{figure}
\begin{center}
     \includegraphics[width=13cm]{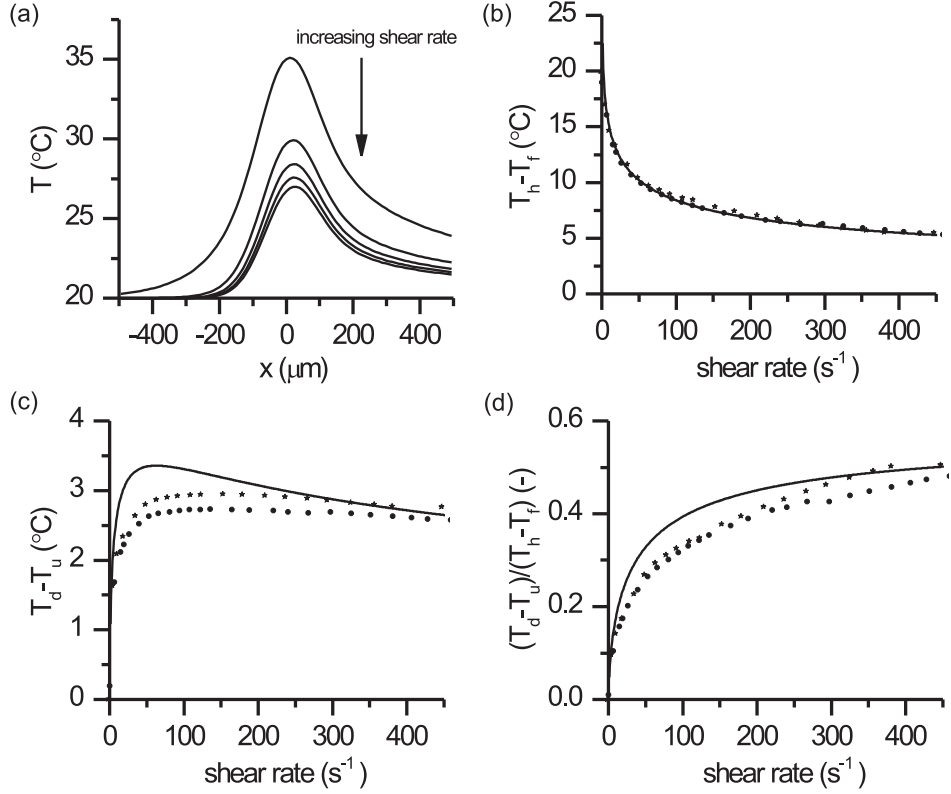}
   \caption{Analytical (---) and experimental ($\bullet\bullet\bullet$, $\star\star\star$) results in steady flow. (a) Analytical temperature profiles at the sensor surface at $S=$10, 60, 110, 160, 210 s$^{-1}$, (b) response of the thermopile measuring $\Th-\Tf$, (c) $\Td-\Tu$, and (d) the ratio of thermopile outputs.}\label{subfigstat}
    \end{center}
\end{figure}

In figure \ref{subfigstat}\,(\textit{a}), the theoretical temperature profiles over the sensor surface are depicted for different shear rates (hence different $\Pec$-values). At low shear rates (low $\Pec$), the temperature distribution is more symmetric with respect to the heater centre, since in that case conduction is dominating the heat-transfer process.
As the shear rate increases the temperature distribution becomes asymmetric, because more heat is advected downstream, while the overall sensor temperature decreases because of the augmented advective cooling.

The experimental data obtained in steady flow are plotted together with the theoretically predicted sensor output in figure \ref{subfigstat}\,(\textit{b--d}). The results of two separate experiments are shown to give an indication of the data spreading, where each data point represents the average result of 20 s of measurement with a sampling frequency of 20 Hz (\ie 400 samples).
Both the experimental and analytical curves show a steep decline in the relative heater temperature $\Th-\Tf$ at low shear rates, and a more gradual one at higher shear rates. Sensor output $\Td-\Tu$ is in both model and experiment characterized by a steep increase at low shear rates, followed by a maximum and a decline; see figure \ref{subfigstat}\,(\textit{c}). These features were also found by \cite{Lammerink93} and \cite{Nguyen95}. With increasing shear rate, the ($\Td-\Tu$)-temperature difference rises because of augmented advection of heat in downstream direction. At the same time the overall sensor temperature decreases (see also figure \ref{subfigstat}\,(\textit{a})), hence, a maximum in $\Td-\Tu$ is observed. Apart from describing very well the qualitative sensor response, the analytical model also predicts the quantitative data with adequate precision. Maximum deviations between model and experiment ranged from 5\% for the ($\Th-\Tf$)-signal to 27\% for $\Td-\Tu$. Although also measurement inaccuracies may play a role, the discrepancy between theory and experiment is most likely due to the simplified modelling of the substrate: the Gaussian heat-flux distribution is only a rough approximation, since the heat transfer from the substrate to the fluid will be larger upstream than downstream, due to the hot thermal wake. Furthermore, the influence of conduction in the substrate decreases with increasing wall shear-rate \cite[see][]{Tardu05}.

\begin{figure}
\begin{center}
     \includegraphics[width=13cm]{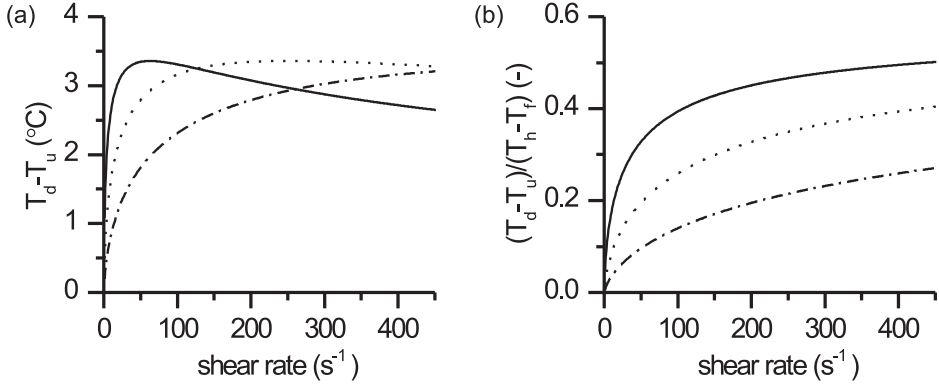}
   \caption{Theoretical results with the original heater width (---), 50\% ($\cdots$), and 25\% (- $\cdot$ -) of the original width for (a) $\Td-\Tu$ and (b) $(\Td-\Tu)/(\Th-\Tf)$.}\label{sensordesign}
    \end{center}
\end{figure}

To obtain an invertible relation between the sensor output and the wall-shear rate or P\'eclet number, the ratio of thermopile outputs, $(\Td-\Tu)/(\Th-\Tf)$, can be used; see figure \ref{subfigstat}\,(\textit{d}). Note that this curve is independent of the thermopile calibration, because the Seebeck coefficient is equal for both thermopiles and vanishes when the ratio of outputs is used. From figure \ref{subfigstat} it appears that the sensor is most sensitive to lower shear rates. The performance of the sensor at higher shear rates, important for the eventual application of the sensor in coronary flow, can be improved by decreasing the heater width $b$, thereby reducing $\Pec$. When the heater width $b=2\sigma$ is decreased, the distance to the heater centre of the thermopile measuring $\Td-\Tu$ must be reduced by an equal amount, to keep the same relative positions. The theoretical results for decreasing the heater width by 50\% and 25\% are shown in figure \ref{sensordesign}. A smaller heater leads to a shift in the maximum temperature difference $\Td-\Tu$, resulting in a more linear relation between the shear rate and the ratio of thermocouple outputs, with a lower sensitivity for lower, and a higher sensitivity for higher shear rates compared to the original response. From figure \ref{sensordesign}\,(\textit{a}) we conclude that the effective heater width has a large influence on the ($\Td-\Tu$)-signal. This could also be an explanation for the discrepancy between theory and experiment shown in figure \ref{subfigstat}\,(\textit{c}); if the effective heater width in the experiment is somewhat smaller than the theoretically used value, this will shift the maximum in the ($\Td-\Tu$)-curve to higher shear rates.  The difficulty here is that the effective heater width will depend on the actual wall-shear rate (\ie the relative influence of conduction in the substrate), making $b$ a function of $S$. The actual effective heater width can therefore only be obtained by solving the conjugate heat-transfer problem.

\begin{figure}
\begin{center}
     \includegraphics[width=13cm]{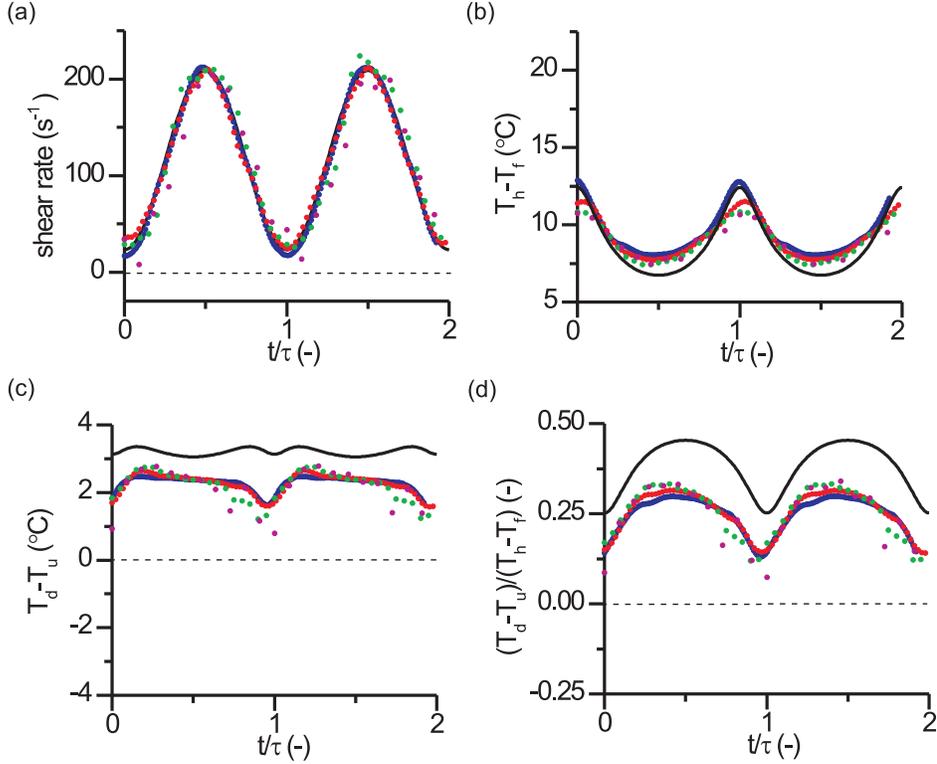}
   \caption{Results in unsteady flow for $\beta=0.8$, with $\tau$ being the period of a flow cycle; analytical (-----) and experimental ($\bullet\bullet\bullet$) curves for \Str=0.01 (blue), \Str=0.03 (red), \Str=0.06 (green), \Str=0.1 (magenta). (a) Shear rate at the sensor surface obtained from the Womersley approximation of the measured flow, (b) response of thermopile measuring $\Th-\Tf$, (c) $\Td-\Tu$, and (d) the ratio of thermopile outputs.}\label{subfiginstat08}
\end{center}
\end{figure}

The sensor response to unsteady flow was investigated experimentally by varying the oscillation frequency, and thereby the Strouhal number, and amplitude in the estimated physiological regime. At each amplitude and frequency, the dynamic sensor response was measured during at least 5 flow cycles; here two periods of each signal are shown. In figure \ref{subfiginstat08} the experimental results for non-reversing shear rates at four \Str-values are depicted, together with the quasi-steady analytical solution. The shear-rate signals calculated from the flow measurements are aligned, to ensure that the phase differences observed in the thermopile signals are due to thermal unsteady effects, and not to phase differences between the flow and the wall-shear rate. Owing to limitations of the pump, the 2 Hz (\Str=0.1) flow signal was not purely sinusoidal, which also shows in the sensor response.

We observe a phase shift and decrease in amplitude with increasing Strouhal number in the ($\Th-\Tf$)-signal. The ($\Td-\Tu$)-thermopile output also shows this phase shift, together with a slight change in the signal shape. As $\Str$ increases, the wall-shear rate oscillations become too fast for the thermal boundary layer to react instantaneously, and the sensor response starts to deviate from its quasi-steady behaviour, and hence from the analytical solution. The deviation between the signals with $\Str=0.01$ and $\Str=0.1$ is larger (17\% in $\Th-\Tf$) during minimum shear rate, when unsteady effects are most important, than during maximum shear rate (6\%), when advection dominates. Not only during minimum shear rate, but during the complete deceleration phase the spreading between the different $\Str$-curves is somewhat larger. To investigate whether flow instabilities in the deceleration phase can explain these deviations, a frequency analysis of the signals was performed. No coherent structures with a fixed frequency were found, so the origin of the deviations remains unclear. Nevertheless, the quasi-steady analytical solution appears to describe the sensor response quite well in the complete experimental range, up to $\Str=0.1$, with again larger quantitative differences in the ($\Td-\Tu$)-signal than the ($\Th-\Tf$)-signal. In the coronary flow regime, with Strouhal numbers of about 0.01 for the first harmonic, a quasi-steady sensor response is therefore expected. In their studies with hot-film anemometers and electrochemical wall-shear probes, respectively, \cite{Clark74} and \cite{Steenhoven91} found the quasi-steady regime to hold for $\Str$ up to 0.2. 

\begin{figure}
\begin{center}
     \includegraphics[width=13cm]{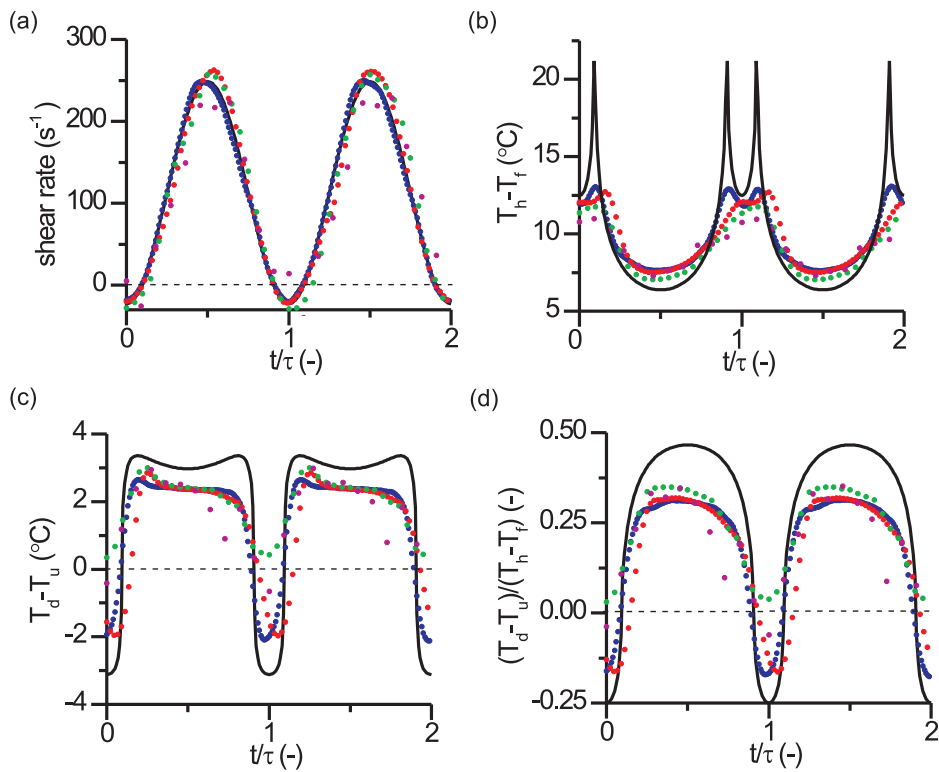}
   \caption{See figure \ref{subfiginstat08}, with $\beta=1.2$.}\label{subfiginstat12}
    \end{center}
\end{figure}

For $\beta=1.2$, larger deviations between the sensor response for $\Str=0.01$ and $\Str=0.1$ have been found during the reversal period and the deceleration phase; see figure \ref{subfiginstat12}. A sign change in $\Td-\Tu$, indicating shear-rate reversal, was clearly observed for the two lowest $\Str$-values. During the reversal period, hot fluid from the thermal wake is carried back over the sensor, which is not taken into account in the analytical model and leads to further deviations from the quasi-steady response. As the shear rate approaches zero, the heat is carried away from the heater only very slowly, leading to large heater temperatures in the quasi-steady analytical solution, while $\Td-\Tu$, and therefore also $(\Td-\Tu)/(\Th-\Tf)$, tend to zero, due to the symmetric influence of conduction. In the experimental data such large relative heater temperatures are never reached because it takes time to heat the fluid, due to its finite thermal diffusivity. Hence, only during the very short period of time where $S$ is close to zero, larger deviations from the quasi-steady solution are observed in the ($\Th-\Tf$)-signals. 

\section{Conclusions}
An analytical model describing the response of a miniature calorimetric sensor to both steady and pulsatile tube flow is developed. In experiments the sensor is subjected to a flow in the expected physiological range to verify the theoretical predictions.
Steady flow analytical and experimental results are in good agreement for the complete range of P\'eclet numbers studied. Hence, our two-dimensional model with the wall-shear rate at the sensor surface as the only flow parameter is sufficient for examining the steady sensor behaviour. Only a simplified model of the substrate in which the sensor is embedded was taken into account, by means of a heat-flux boundary condition. A conjugate approach will lead to a more accurate quantitative prediction of the temperature differences measured, however, our model has the advantage of a simple representation of the substrate, and still leads to an acceptable description of the sensor response.

The quasi-steady analytical model predicts the sensor behaviour in non-reversing pulsatile flow with Strouhal numbers up to 0.1 quite well. Based on the experimental results, we conclude that the sensor response to coronary flow will be quasi-steady, except during the (short) periods of shear-rate reversal. The analytical model can therefore be used to optimize the sensor design for coronary flow measurements, as demonstrated in figure \ref{sensordesign}.

\begin{acknowledgements}
This research was financially supported by the Dutch Technology Foundation STW, project SmartSiP 10046, Philips Research, and St. Jude Medical.
\end{acknowledgements}

\end{document}